# Carbon Dimer in Silicon Cage: A Class of Highly Stable Silicon Carbide Clusters

M. N. Huda and A. K. Ray*
Department of Physics, The University of Texas at Arlington, Arlington, Texas 76019

## Abstract

A class of silicon carbide cage clusters with two carbon atoms *inside* the silicon cage and with *high* stabilities are presented. The theoretical formalism used is Hartree-Fock theory followed by second order many body perturbation theory to account for correlation effects and geometry optimizations at the second order perturbation theory level are performed without any symmetry constraints. The smallest "cage" is found to be a silicon cube with the carbon dimer inside the cube. Based on the simultaneous criteria of high binding energy, high vertical ionization potential, high homo-lumo gap, and low vertical electron affinity, $Si_{14}C_2$, with a close fullerene like structure, is predicted to be a particularly stable cluster both at all-electron and at pseudopotential level of calculations. The C-C bond lengths and the HOMO-LUMO gaps of the clusters are both found to oscillate with cluster size.

PACS : 36.40*; 61.48+c

Clusters are distinctly different from their bulk state and exhibit many specific properties, which distinguishes their studies as a completely different branch of science named "Cluster Science". Large surface to volume ratio and quantum effects resulting from small dimensions are usually prominent in clusters and ideas like 'super-atoms', 'magic numbers' or 'fission' in clusters [1 – 4] have prompted a wide class of scientists to study this 'relatively' new area of the physical sciences. Growing interests in the stabilities of small clusters and the evolution of bulk properties from cluster properties are also due to the emergence of new areas of research called nanoscience and nanotechnology and the resulting potentials in industrial applications. In the study of clusters, cage-like compact structures are particularly important for two reasons; first, they can be used as building blocks of more stable materials. Secondly, the hollow space inside a cage can be used to dope different suitable atoms to yield a wide variety of atomically engineered materials with specific properties. For example well-controlled nanostructures with varying HOMO-LUMO gaps and desired conduction properties can be achieved by controlled doping of atoms in $C_{60}$ clusters [5]. The spin property of the doped atom inside the cage can be used as the smallest memory devices for future quantum computers; for instance, tungsten in $Si_{12}$ clusters is quantum mechanically isolated from outside so it can preserve its spin state [6].

Silicon is one of the most important semi-conducting elements with widespread applications in industry and silicon clusters, preferring $sp^3$ hybridization, have been studied extensively in the literature. Different theoretical methods based on Hartree-Fock (HF) and density functional theories (DFT) [7-9] have been used to determine the ground state structures of bare Si clusters. Still much debate exists on the structures of medium sized Si clusters, because a large number of topologically different structures are possible as the number of atoms in the cluster increases [10]. Discovery of the magically stable $C_{60}$ fullerene cage have prompted scientists to study fullerene-like silicon structures and it was found that the $Si_{60}$ has a distorted fullerene-cage-like structure [11]. Attempt also has been made to replace carbon atoms by Si atoms in $C_{60}$, resulting in a distorted structure [12]. In carbon clusters, preferring $sp^2$ hybridization, fullerene like structures are found experimentally in as small as $C_{20}$ clusters [13]. In $Si_n$ clusters, starting from n=19, ground state structures drastically change from prolate

---

*email: akr@exchange.uta.edu

assemblies of $Si_9$ subunits to more compact cage like geometries [10].

It has been pointed out recently, primarily based on DFT studies, that highly stable small silicon cage clusters are possible if transition *metal* atoms are encapsulated in the cage [14]. The combinations of silicon and carbon atoms in a cluster have generated a number of studies. These studies, ranging in areas from cluster science to astrophysics have primarily concentrated on clusters rich in carbon atoms [15]. However, to the best of our knowledge, *Si rich cage type* silicon carbide clusters have not been studied in detail so far. We show below that *carbon dimers* trapped into medium size silicon clusters produces structures *comparable in stability* to metal encapsulated silicon cage clusters. For this purpose, we have carried out *ab initio* Hartree-Fock based second order Møller Plesset perturbation theory calculations to study the electronic and geometric structures of $Si_nC_2$ (n = 8-14) clusters. We have recently applied this theory to study silver clusters and predicted $Ag_8$ to be a magic cluster [16]. Both all electron and pseudopotential basis sets have been used. For the all electron calculations, a 6-311G** [8] basis set and for the other, effective core potentials with Hay-Wadt basis sets have been used [17]. Geometry optimizations have been carried out at the second order perturbation theory level and all computations have been performed using the *Gaussian 98* suite of programs [18].

In table 1, we have listed the binding energy (BE), vertical ionization potential (VIP), vertical electron affinity (VEA), highest occupied molecular orbital – lowest unoccupied molecular orbital (HOMO-LUMO) gap, and the C-C bond length for all the clusters. Here the binding energies are calculated at the separated atom limit, with the atoms in their respective ground states. Figure 1 shows the optimized cluster structures obtained at the MP2. For each cluster, we have optimized several *possible* structures; only the most stable structures are shown here [19]. The smallest structure is a cube like structure with silicon atoms at the corner of the cube and two carbon atoms inside the cube. The width of the cube is 2.58 Å, and the C–Si bond length is 1.93 Å with 6-311G** basis set. The height of the cube is 0.15 Å longer than its width. The binding energy of this cluster was found to be 37.44eV. The C-C bond length is 1.43 A, to be compared with the free carbon dimer bond length of 1.27 A. Thus, the C-C bond length stretches by 0.16 Å in the presence of the silicon atoms. From Mulliken charge analysis, we find that the carbon atoms are negatively charged with −1.06e ( −1.75e with Hay-Wadt basis set), while all the Si atoms are slightly positively charged. The bonding is thus partially ionic. As regards stability is concerned, the cluster also has a high VIP of 7.72eV, high HOMO-LUMO gap of 6.64eV and a rather low VEA of 1.19eV. We note here that in metal doped fullerenes, metal atom act as an electron donor, whereas in the present case doped carbon atoms acquired negative charges. Also, the electronic and geometric structure quantities are, to an extent, method and basis-set dependent. For example, for $Si_8C_2$, we carried out optimizations with two other basis sets, namely STO-3G and 3-21G. The homo-lumo gap was 6.590eV for the minimal basis set and 6.004eV for the 3-21G set. The primary conclusions about this unique set of clusters are however expected to hold true, irrespective of the methodology used. $Si_9C_2$ cluster is a cage made up with a rectangle and a pentagon of Si atoms, with $C_2$ inside the cage. The starting geometry of this cluster before optimization was a capped cubic structure, *i.e.*, a Si cap was added to the optimized $Si_8C_2$ cluster. The binding energy for this cluster is 42.46eV, with a HOMO-LUMO gap of 5.89eV. The vertical ionization potential is 7.88eV and the VEA is 1.34eV. The height of the cage is 2.56 Å and C-C bond length is 1.48 Å. This cluster is not as symmetrical as the previous one, $Si_8C_2$; as a consequence, it has a small dipole moment and nonzero non-diagonal quadrupole moments. Similar to the $Si_8C_2$ cluster, the carbon atoms are negatively charged. The average Si-C bond length is almost the same as in the cubic cluster.

$Si_{10}C_2$ cluster, after optimization, assumes the structure of two pentagons of Si atoms back to back with the two carbon atoms parallel to the pentagonal face. The C-C bond length of 1.52A is slightly greater than the previous cluster. The average C-Si bond length is 1.985 Å, which is higher than the previous two clusters. From the structural point of view, this cluster is more compact than the $Si_9C_2$ clusters. The vertical ionization potential is 8.47eV and the vertical electron affinity is 1.81eV, with a HOMO-LUMO gap of 4.97eV. For $Si_{11}C_2$, the structure is not exactly a cage type structure. It is a tricapped twisted cubic Si-frame with the two carbon atoms at the middle joining the two faces; Si atoms in the two twisted opposite capped surfaces are too separated to make it a closed structure. One of the silicon atoms forms a bridge between these two separated capped faces on one side of the stretched cube. The C-C bond length is 1.46 Å, and the Si-C average bond length is almost the same as the first two structures. Binding energy of this cluster is 49.01eV, with a HOMO-LUMO gap of 6.69eV. Vertical ionization potential for this cluster is 7.40eV. In spite of the fact that this cluster is more prolate-type, the Mulliken charge distribution shows the same pattern as the smaller clusters.

$Si_{12}$-cage cluster has been studied extensively in the literature, because of the formation of magic clusters with metal atoms, such as W or Cr, inside. However, magicity is not well defined for the clusters whose constituents are non-metallic atoms, as pointed out by Khanna *et al.* [14]. This is mainly because there is no jellium like model for covalently bonded non-metallic clusters, so the shell closing at 2, 8, 20, … of valence electrons is not theoretically justified here. For $Si_{12}C_2$, with Hay-Wadt basis set, two competing cage clusters are the four Si capped cubic cage and distorted hexagonal cage, with the distorted cage being slightly lower in energy. For the cubic cage the two carbon atoms are completely within the cage with a C-C bond length of 1.54 Å. For the distorted hexagonal cage, the two carbon atoms are also inside the cage but parallel to the base of the cluster. The overall charge distribution pattern is the same as the other smaller clusters. One point is to note that between these two structures, in the hexagonal cage the carbon atoms are less negative. Optimization with the all electron basis set also found the distorted hexagonal cage as the ground state. Vertical ionization potential is 7.70eV for this cage and VEA is 2.91eV. The C-C bond length in this case is 1.54 Å. Binding energy for this cluster is 42.42eV and 53.62eV at the pseudopotential and all-electron basis sets, respectively. This binding energy is comparable to the binding energy 43.73eV of tungsten incorporated $Si_{12}$ cage, $Si_{12}W$, at the pseudopotential level.

$Si_{13}C_2$ and $Si_{14}C_2$ are formed by putting Si-cap on the hexagonal cage of $Si_{12}C_2$. The hexagonal cage is distorted, but in both of the two cases two carbon atoms are inside the cage. The $Si_{14}C_2$ cluster looks like a distorted sphere made with 12 irregular rectangles, where carbon atoms are not exactly in the center of the sphere. The binding energies for these two clusters are 57.04eV and 62.24eV, respectively. HOMO-LUMO gap for the bigger cluster is about 1.3eV higher. The vertical ionization potential for the $Si_{14}C_2$ is 8.58eV, which is very high. However the main difference between these two clusters and the previous set is in the charge distribution. Along with the two negatively charged carbon atoms, some other Si atoms also became slightly negative. C-C bond lengths for 13 and 14 atom Si clusters are 1.45 Å and 1.52 Å respectively.

In figures 2 and 3, we have plotted the binding energies per atom, vertical ionization potentials, vertical electron affinities, homo-lumo gaps, and the carbon dimer bond lengths versus the number of silicon atoms in the cluster, with both the all electron and the pseudopotential basis sets. At the all electron basis set level, $Si_{14}C_2$ has the highest binding energy but at the pseudopotential level, $Si_9C_2$ has the highest binding energy. The general pattern is similar, with the all electron binding energies being consistently higher than the pseudopotential binding energies. We also note that *all* the VIPs are rather high and *all* the VEAs are rather low. In addition, for all the clusters, the HOMO-LUMO gaps are also very high. Thus, we can infer that the reactivity for this class of clusters is not very high. In the plot for VIP, after an initial peak at $Si_9C_2$, it decreases and then increases for the highest peak at $Si_{14}C_2$ with the Hay-Wadt basis set, whereas with the 6-311G** basis set, initial peak is at $Si_{10}C_2$ followed by the highest pack at $Si_{13}C_2$. However, $Si_{14}C_2$ has also considerably high VIP. It also has the lowest VEA of 0.05eV at the all-electron basis set and 0.08eV at the Hay-Wadt basis set level. The very high IP and lowest EA of $Si_{14}C_2$ indicate that this cluster may be a candidate for a highly stable cluster. Only for this cluster, we have a distorted spherical shape, implying that the competition between $sp^2$ and $sp^3$ hybridizations is rather balanced here. Average coordination number per atom of this cluster is 3.87. We also note that its HOMO-LUMO gap is higher than the neighboring clusters. In the plot showing the variations of the HOMO-LUMO gap with the cluster size, both plots with the two basis sets indicate a peak at $Si_{11}C_2$. This cluster also has considerably high IP but its electron affinity is also high compared to most of the other clusters. This cluster is not exactly a cage-type, but more open type.

As a general feature, we also note, as mentioned before, carbon atoms gain charge whereas silicon atoms lose charge. This is also to be expected from electronegativity considerations. In the plot for the C-C bond length versus the number of Si atoms in a cluster, an oscillatory pattern is observed and some correlation is observed between the C-C bond length and the HOMO-LUMO gap. At the level of both basis sets, the local minima for the C-C bond length for $Si_{11}C_2$ corresponds to a local maxima for the HOMO-LUMO gap. Up to $Si_{13}C_2$, the curves of C-C bond lengths are similar. Stronger overlapping of the C-C orbitals contributes to the stability of the clusters, and C-C bond length directly influences the HOMO-LUMO gap. We note here, as was pointed out by Grev and Schaefer [20], that the Si-C $\pi$-bond energy is 70% of the analogous C-C $\pi$-bond energy, whereas for $\sigma$-bond both the C-C and Si-C bonds are of the same strength. Typically, silicon atoms prefer $\sigma$ bonding and carbon atoms, $\pi$ bonding. As the C-C distance increases, the likelihood of $\pi$ bonding between them decreases. Usually higher C-C bond length means higher height of the cage; this, in turn, implies weaker overlapping of Si atoms situated in the upper and lower surfaces of the cage, so the $\sigma$ bonding became weaker between these atoms. There are some Si atoms in the cage which are threefold coordinated, *e.g.*, in $Si_{13}C_2$ there are six three fold coordinated atoms. This implies two possibilities: firstly in the case of *$sp^3$* bonding between the atoms, there should be a dangling bond associated with the three-fold coordinated atoms. Second possibility is that if the bonding is *$sp^2$* then either there would be some double bonding between the silicon in the cage, or there would be again dangling bonds. The effect of the dangling bonds is to destabilize the clusters. Though the structures are distorted from their perfectly symmetrical situation, the structures did stabilize. So the bonding here is a mixture of *$sp^2$* and *$sp^3$* hybridization to minimize the dangling bond effect, which is well known in case for fullerenes. As a final comment, as these are Si rich clusters, *$sp^3$* hybridization dominates, and this may be responsible for the distortion seen in the clusters from the symmetry. However the reasons behind the oscillatory behaviors in HOMO-LUMO gap and C-C bonding needs more study.

In summary, we studied here a class of silicon carbide cage cluster with higher stability where the two carbon atoms sit within the silicon cage. Second order many body perturbation theory was used to take into account the electron correlations and full geometry optimization was done without any symmetry constraint. The smallest cage was found to be a silicon cube with carbon dimer inside the cube. It was found that the $Si_{14}C_2$ is almost a spherical structure with very high VIP, lowest VEA and higher HOMO-LUMO gap. In all the clusters studied here, doped carbon atoms become negatively charged, unlike any other metal doped fullerenes. It was also found that the C-C bond length has direct influence on the HOMO-LUMO gap of the clusters. They both oscillate with respect to cluster size.

Finally, the authors gratefully acknowledge partial support from the Welch Foundation, Houston, Texas (Grant No. Y-1525).

Table 1. Binding energies, vertical ionization potentials, HOMO-LUMO gaps, vertical electron affinities (all in eV) and the C-C bond lengths (in Å). Numbers in parenthesis are for the Hay-Wadt basis set.

| Clusters | B.E. in eV | VIP in eV | VEA in eV | HOMO-LUMO gap in eV | C-C bond length in Å |
|---|---|---|---|---|---|
| $Si_8C_2$ | 37.44 (30.15) | 7.72 (7.85) | 1.19 (1.22) | 6.64 (6.17) | 1.43 (1.45) |
| $Si_9C_2$ | 42.46 (33.55) | 7.88 (8.27) | 1.34 (1.28) | 5.89 (5.81) | 1.48 (1.50) |
| $Si_{10}C_2$ | 45.96 (36.24) | 8.47 (7.54) | 1.81 (0.86) | 4.97 (5.52) | 1.52 (1.52) |
| $Si_{11}C_2$ | 49.01 (38.74) | 7.40 (7.31) | 1.73 (1.49) | 6.69 (6.17) | 1.44 (1.48) |
| $Si_{12}C_2$ | 53.62 (42.42) | 7.70 (7.27) | 1.24 (1.25) | 5.98 (5.53) | 1.54 (1.55) |
| $Si_{13}C_2$ | 57.04 (45.45) | 9.02 (7.14) | 1.23 (0.62) | 4.13 (4.21) | 1.45 (1.52) |
| $Si_{14}C_2$ | 62.24 (48.16) | 8.58 (9.23) | 0.05 (0.08) | 5.43 (5.21) | 1.52 (1.49) |

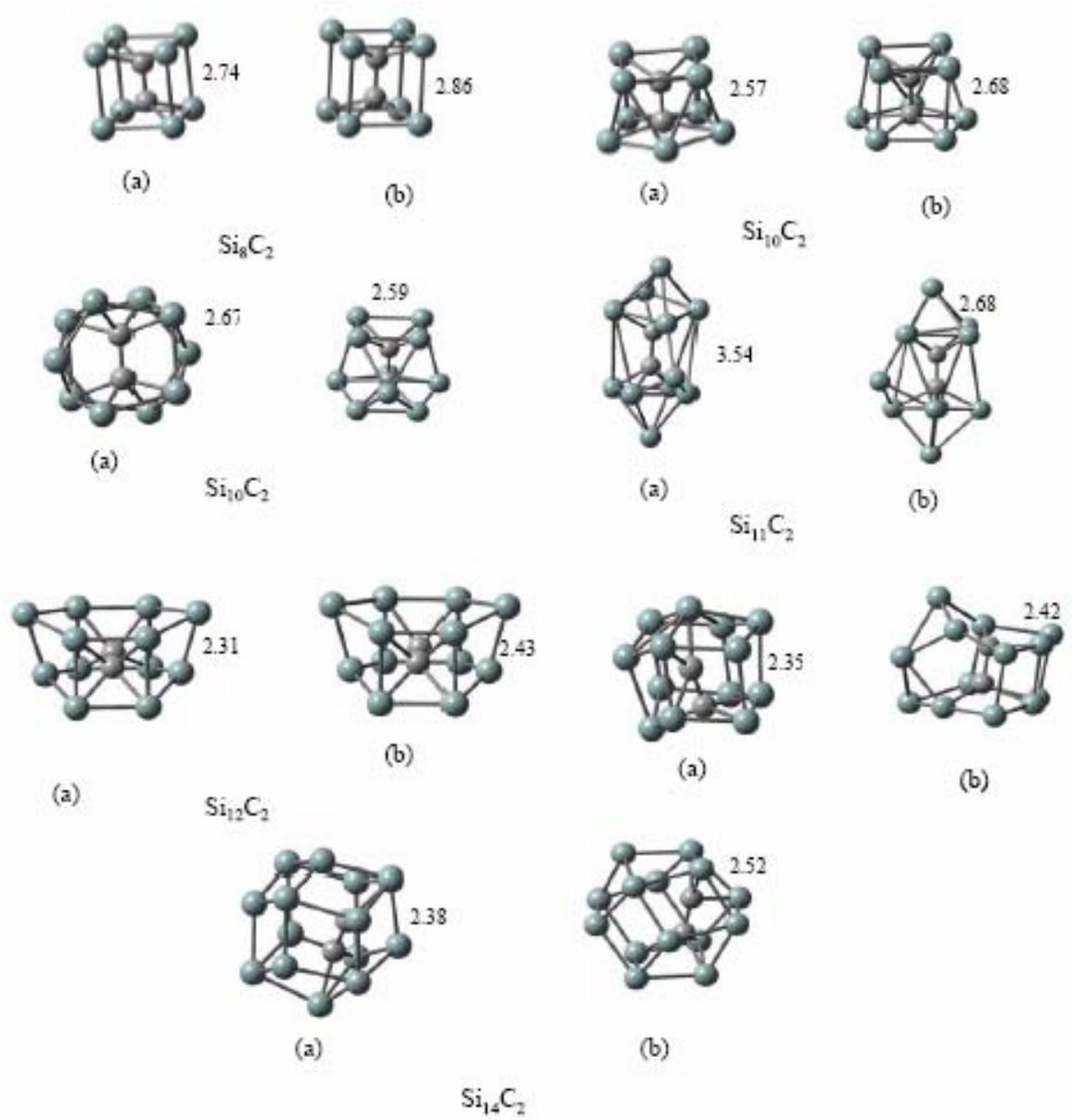

Figure 1. Optimized cluster structures (a) by 6-311G** basis set (b) by Hay-Wadt basis set.

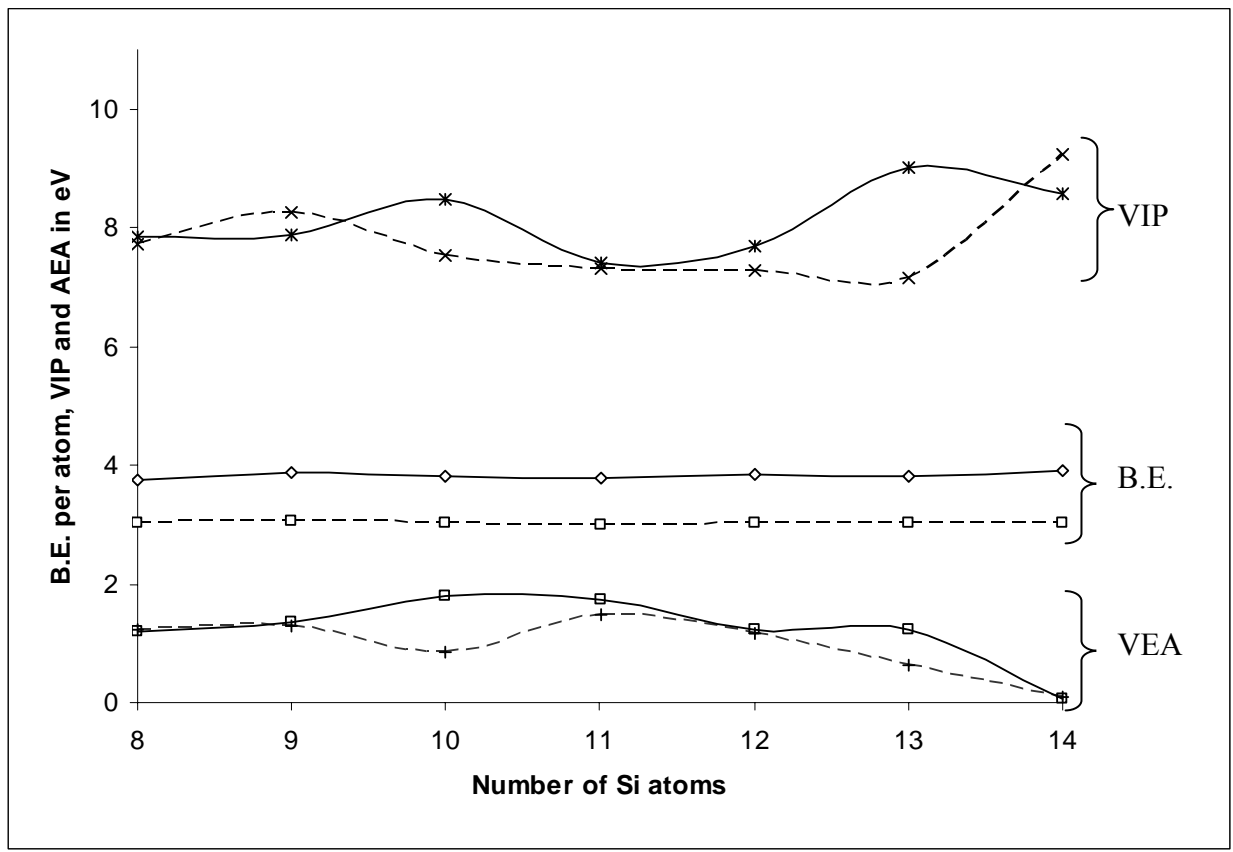

Figure 2. Binding energies per atom, vertical ionization potentials and electron affinities versus the number of silicon atoms with 6-311G** basis set (solid line) and Hay-Wadt basis set (broken line).

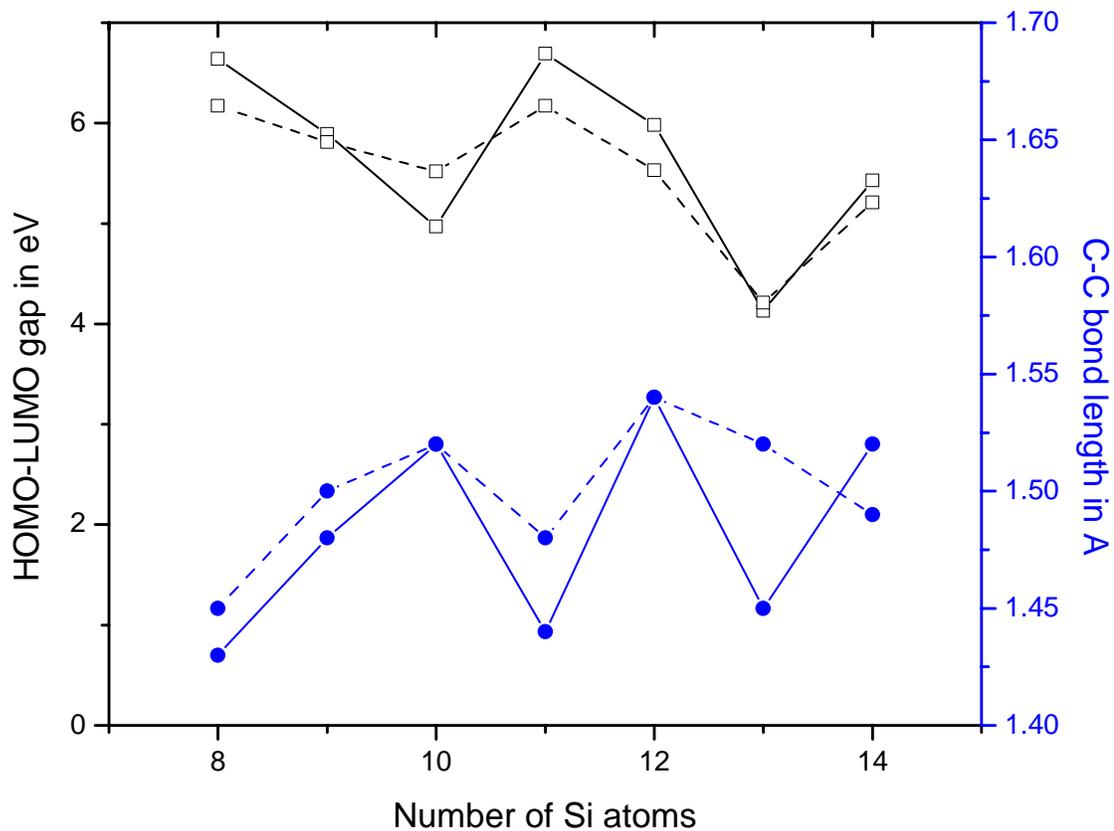

Figure 3. HOMO-LUMO gaps (in eV) and C-C bond lengths (in Å) versus the number of silicon atoms with 6-311G** basis set (solid line) and Hay-Wadt basis set (broken line).